\documentclass[twocolumn,showpacs,preprintnumbers,amsmath,amssymb]{revtex4}
\usepackage{graphicx}
\usepackage{dcolumn}
\usepackage{bm}

\begin{document}

\title{Collective excitations of Dirac electrons in a graphene layer
with spin-orbit interaction}
\author{X.F. Wang and Tapash Chakraborty}
\affiliation{Department of Physics and Astronomy, The University of
Manitoba, Winnipeg, Canada, R3T 2N2}

\begin{abstract}
Coulomb screening and excitation spectra of electrons in a graphene
layer with spin-orbit interaction (SOI) is studied in the random phase
approximation. The SOI opens a gap between the valence and conduction
bands of the semi-metal Dirac system and reshapes the bottom and top of
the bands. As a result, we have observed a dramatic change in the
long-wavelength dielectric function of the system. An undamped plasmon
mode emerges from the inter-band electron-hole excitation continuum edge
and vanishes or merges with a Landau damped mode on the edge of the
intra-band electron-hole continuum. The characteristic collective
excitation of the Dirac gas is recovered in a system with a high carrier
density.
\end{abstract}
\maketitle

Since the recent discovery of quantum Hall effect in graphene
\cite{novo,zhan,qhe}, there has been an upsurge of interest in
understanding its electronic properties. Graphene has a promising
potential for nanoscale device applications \cite{today} and is
also very interesting physically because of its unusual Dirac-Weyl
type band-structure near the Fermi level
\cite{wall,ando1,hald,kane,sini,khve,pere}. As a single layer of graphite
or an unrolled single-walled nanotube, the energy band structure of graphene
is well known just as for the graphite or for a nanotube \cite{wall,ando1,kane}.
The past few years have also evidenced increased activities in the role of
spin-orbit interaction (SOI) in nanostructures because it introduces
many unusual features in these systems. The SOI is also found to exhibit
interesting effects such as the spin Hall effect in graphene \cite{kane}
but overall its effect in graphene is as yet, unclear \cite{ando2}.
Further, the electron-electron interaction is important to understand
the behavior of electrons in graphene and has been studied by many
authors in graphite-based structures but without the SOI included
\cite{gonz,alva,vafe,ando2}. In this letter, we explore the effect of SOI on the
electron-electron interaction and the characteristic excitations in a
graphene layer.

Graphene has a honeycomb lattice of carbon atoms with two sublattices.
Its energy band can be calculated by the tight-binding model \cite{wall,ando1}
and an intrinsic graphene is a semimetal with the Fermi energy located at
the inequivalent $K$ and $K'$ points at opposite corners of its
hexagonal Brillouin zone \cite{wall}. In the effective mass approximation,
the Hamiltonian of electrons near the Fermi energy is expressed by a
$8\times 8$ matrix. Since the SOI due to the atomic potential only
mixes the states corresponding to the two sublattices, we can reduce
the matrix to four independent $2\times 2$ blocks. In the
representation of the two sublattices, the Hamiltonian of a spin-up
electron near the $K$ point of the reciprocal space reads \cite{kane,sini}
\begin{equation}
H =
v\bm{p}\cdot\bm{\sigma}+\Delta_{so}\sigma_z
=\left[
\begin{array}{cc}
 \Delta_{so}&-i\hbar v \nabla^-\\
-i\hbar v \nabla^+ & -\Delta_{so}
\end{array}
\right],
\end{equation}
with $\bm{\sigma}=(\sigma_x,\sigma_y,\sigma_z)$ the Pauli matrices
in the pseudospin space of the two sublattices and $\nabla^{\pm}= \partial/\partial
x\pm i\partial/\partial y$. Here $\Delta_{so}$ is the strength of the
spin-orbit interaction (SOI) and $v=10^6$ m/s is the `light' velocity
of the Dirac electron gas.  The eigenfunctions are, $\Psi_{\bm{k}}^+(
\bm{r})=e^{i\bm{k}\cdot\bm{r}} {\small\left( \array{c}
\cos(\alpha_{\bm{k}}/2) \\
e^{i\phi_{\bm{k}}}\sin(\alpha_{\bm{k}}/2)
\endarray
\right)}$ for the state $|\bm{k},+\rangle$ in the conduction band of energy
$E_{\bm{k}}^+=\sqrt{\Delta_{so}^2+\hbar^2v^2k^2}$ while
$\Psi_{\bm{k}}^-(\bm{r})=e^{i\bm{k}\cdot\bm{r}}
{\small\left(
\array{c}
\sin(\alpha_{\bm{k}}/2) \\
-e^{i\phi_{\bm{k}}}\cos(\alpha_{\bm{k}}/2)
\endarray
\right)}$
for the state $|\bm{k},-\rangle$ in the valence band
$E_{\bm{k}}^-=-\sqrt{\Delta_{so}^2+\hbar^2v^2k^2}$ with $\tan\phi_{\bm{k}}
=k_y/k_x$, $\tan\alpha_{\bm{k}}=\hbar v k/\Delta_{so}$, and $k=\sqrt{k_x^2+
k_y^2}$. For a bare Coulomb scattering of two electrons at states
$|\bm{k},\lambda\rangle$ and $|\bm{p},\lambda_1\rangle$ into states
$|\bm{k}+\bm{q},\lambda'\rangle$ and $|\bm{p}-\bm{q},\lambda'_1\rangle$
respectively, the interaction matrix elements are
\begin{equation}
v_{\bm{k},\bm{p}}^{\lambda,\lambda',\lambda_1,\lambda'_1}
=g_{\bm{k}}^{\lambda,\lambda'}(\bm{q})v_0(q)
g_{\bm{p}}^{\lambda_1,\lambda'_1}(-\bm{q}).
\label{coulomb}
\end{equation}
Here $v_0=e^2/(2\epsilon_0\epsilon_i q)$ is the two-dimensional
Coulomb interaction (in Fourier space) with the high-frequency dielectric
constant $\epsilon_i=1$ \cite{pere},
$g_{\bm{k}}^{\lambda,\lambda'}(\bm{q})$ is the interaction vertex,
and the index $\lambda=\pm$ denotes the two bands.
It is straightforward to show that the RPA dressed interaction matrix
elements have the same form as the bare interaction matrix elements,
i.e. Eq.~(\ref{coulomb}) \cite{vint}. As a result, the dielectric matrix is
expressed as a unit matrix multiplied by a dielectric function
\begin{equation}
\hat{\epsilon}(q,\omega)=1-v_0(q)\hat{\Pi}({\bf q},\omega)
\label{dielsg}
\end{equation}
with the electron-hole propagator
\begin{equation}
\hat{\Pi}({\bf q},\omega) =4\sum_{\lambda, \lambda', \bm{k}}
|g_{\bm{k}}^{\lambda,\lambda'}(\bm{q})|^2
\frac{f[E^{\lambda'}_{\bm{k}+\bm{q}}]-f[E^\lambda_{\bm{k}}]}
{\hbar\omega+E^{\lambda'}_{\bm{k}+\bm{q}}-E^\lambda_{\bm{k}}+i\delta}.
\label{propagator}
\end{equation}
The factor 4 comes from the degenerate two spins and two valleys at $K$ and
$K'$; and the vertex factor reads
$|g^{\lambda,\lambda'}_{\bm{k}}(\bm{q})|^2=
[1+\lambda\lambda'\cos\alpha_{\bm{k}+\bm{q}}\cos\alpha_{\bm{k}}
+\lambda\lambda'\sin\alpha_{\bm{k}+\bm{q}}\sin\alpha_{\bm{k}} (k+q
\cos\theta)/|\bm{k}+\bm{q}|]/2$ with $\theta$ being the angle
between $\bm{k}$ and $\bm{q}$. Since the intra-band
backward scattering at $\bm{q}=2\bm{k}$
 and the inter-band
vertical transition at $\bm{q}=0$
 are not allowed under Coulomb interaction in the
system, we have $|g^{\lambda,-\lambda}_{\bm{k}}(0)|^2=|g^{\lambda,\lambda}_{\bm{k}}(2\bm{k})|^2=0$. The collective excitation spectrum is obtained by finding
the zeros of the real part of the dielectric function $\epsilon_r$.
For convenience we denote each zero as a plasmon mode which may
differ from the convention used in other places where some Landau
damped modes are not counted since they do not have poles for
$\hat{\epsilon}^{-1}$.

\begin{figure}
\begin{center}
\begin{picture}(450,160)
\put(-20,-30){\includegraphics{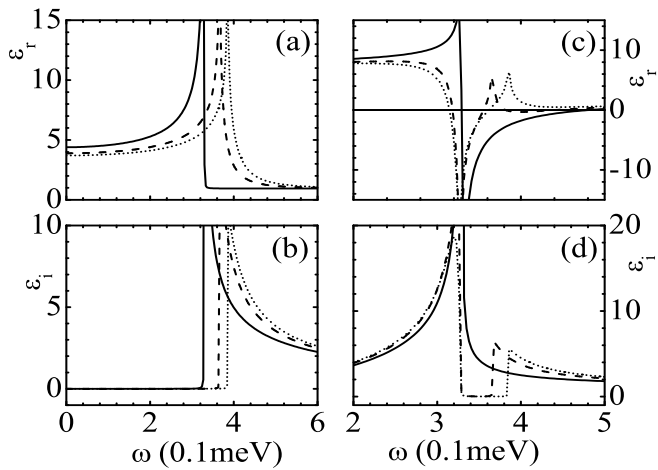}}
\end{picture}
\vspace{0cm} \protect\caption{Real ($\epsilon_r$) and imaginary
($\epsilon_i$) parts of the dielectric function vs $\omega$ at $T=0$
(left panels) and $T=2$ K (right panels) for an intrinsic graphene
($E_F=0$) and the SOI strength $\Delta_{so}=0$ (solid), 0.08 meV
(dashed), and 0.1 meV (dotted). The wave vector is $q=0.05 \times
10^5$ cm$^{-1}$.} \label{fig:fig1}
\end{center}
\end{figure}

First we explore the properties of an intrinsic graphene where the
net electron density is equal to zero and $E_F=0$. In
Fig.~\ref{fig:fig1} (a) and (b), we show the real ($\epsilon_r$) and
imaginary ($\epsilon_i$) parts of the dielectric function at
$T=0$ for different SOI strengths \cite{ando1,kane}
$\Delta_{so}=0$ (solid), 0.08 (dashed), and 0.1 meV (dotted) at a
wave vector $q=0.05 \times 10^5$ cm$^{-1}$. In the absence of the
SOI, i.e. for $\Delta_{so}=0$, the zero temperature electron-hole
propagator is given by $\hat{\Pi}_0({\bf
q},\omega)=-q^2/[4\sqrt{v^2q^2-\omega^2}],$ which has been previously
obtained via the renormalization group theory \cite{gonz,khve}. We
have $\epsilon_r=1$ (for $\omega > vq$) above the inter-band
electron-hole continuum (EHC) edge and $\epsilon_i=0$ (for $\omega <
vq$) below it since only the interband transition is allowed for
electrons. With an increasing $\Delta_{so}$, the peaks of $\epsilon_r$ and
$\epsilon_i$, which are located at the edge of the inter-band EHC,
$\omega_H=2\sqrt{\Delta_{so}^2+\hbar^2v^2q^2/4}$, shift to higher
energies. At the same time, for $\omega > \omega_H$, $\epsilon_r$
increases continuously with $\Delta_{so}$. In the intrinsic graphene
there is no plasmon mode at the zero temperature.

\begin{figure}
\begin{center}
\begin{picture}(220,220)
\put(-15,25){\includegraphics{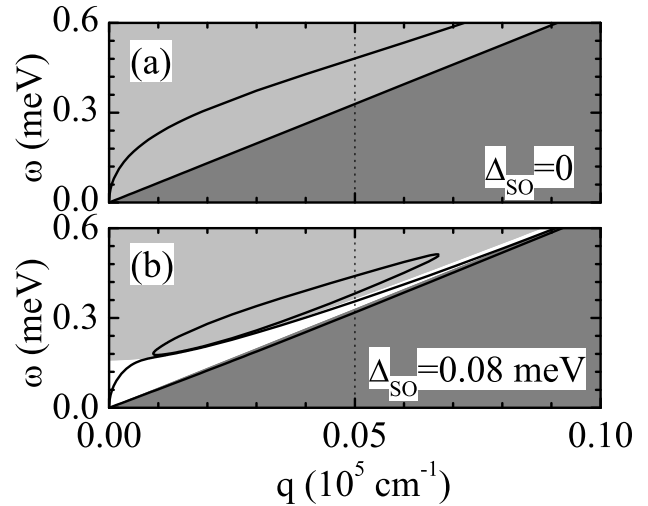}}
\end{picture}
\vspace{-1.4cm} \protect\caption{Plasmon spectrum (thick curves) of
an electron gas in an intrinsic graphene ($E_F=0$) at
$T=2$ K with $\Delta_{so}=0$ in (a) and 0.08 meV in (b). Intra-
(dark shaded) and inter- (light shaded) band single-particle
continuums are also shown. The vertical dotted lines indicate the
$q$ values for which the dielectric function is shown in
Fig.~\ref{fig:fig1} (c).} \label{fig:fig2}
\end{center}
\end{figure}

At a finite temperature, the intra-band transition is allowed and contributes 
to the electron-hole propagator of Eq.~(\ref{propagator}) while the inter-band 
contribution is reduced from that at zero temperature because of the electronic 
occupation of the conduction band. For $\Delta_{so}=0$ where $\omega_L =\omega_H$, 
a $\epsilon_r$ dip at the intra-band EHC edge $\omega_L = vq$ in the high energy 
side is formed as shown by the solid curve in Fig.~\ref{fig:fig1} (c) and two roots 
of the equation $\epsilon_r=0$ or two plasmon modes may appear, one at $\omega=vq$ 
while the other has a higher energy. In Fig.~\ref{fig:fig1} (d), we observe a finite 
value of $\epsilon_i$ for $\omega < \omega_L$ due to the intra-band transition and a 
reduced $\epsilon_i$ for $\omega > \omega_L$ corresponding to the weakening of 
inter-band scattering as the temperature increases. As a result, the plasmon mode 
at $vq$ is strongly damped while the other is weakly damped. With increasing
$\Delta_{so}$, the $\epsilon_r$ peak due to inter-band transitions shifts with 
$\omega_H$ to a higher energy while the $\epsilon_r$ dip of intra-band transitions 
stays with $\omega_L$. Since the $\epsilon_r$ peak has a lower energy than the
$\epsilon_r$ dip initially at $\Delta_{so}=0$, the peak and the dip
will merge at first and split again. As a result, the $\epsilon_r$ curve is deformed 
in such a way that two extra zeros of $\epsilon_r$ or two new plasmon modes emerge 
at a finite $\Delta_{so}$. If $\Delta_{so}$ is further increased the $\epsilon_r$ peak 
finally disappears as the gap between the $\epsilon_r$ dip and peak becomes wider and the
inter-band interaction diminishes. Corresponding to the separation of the
$\epsilon_r$ peak from its dip, the $\epsilon_i$ curve develops a gap
between $\omega_L$ and $\omega_H$ [Fig.~\ref{fig:fig1} (d)].

The $\omega$ versus $q$ spectrum of the plasmon modes at $T=2$ K is
plotted by thick solid curves for $\Delta_{so}=0$ in Fig.~\ref{fig:fig2} (a) and 
$\Delta_{so}=0.08$ meV in Fig.~\ref{fig:fig2} (b). We observe one strongly Landau-damped
plasmon mode at $\omega=vq$ and a weakly damped one in the inter-band EHC with an 
approximate dispersion of $\omega \propto \sqrt{q}$ near $q=0$ in the $\Delta_{so}=0$ 
case. A finite $\Delta_{so}$ separates $\omega_H$ from $\omega_L$ and opens a gap 
between the intra- and inter-band EHC's. Near $q=0$ the former weakly damped plasmon 
mode becomes undamped because it is now located in the gap. Due to a strong inter-band 
single-particle transition at the edge of the inter-band EHC, the dispersion curve of this 
mode is squeezed to a lower energy when it approaches the inter-band EHC edge and then 
splits into three plasmon modes, i.e., two new plasmon modes emerge near the 
inter-band EHC edge. One of the modes remain undamped in the EHC gap (or the $\epsilon_i$ 
gap) while the other two are located inside the inter-band EHC. The latter two modes merge 
and disappear at $q$ near $ 0.07 \times 10^5$ cm$^{-1}$ in Fig.~\ref{fig:fig2} (b). The
undamped mode survives in larger wavevectors until it merges with the
strongly damped mode at the intra-band EHC edge.

As explained in Fig.~\ref{fig:fig1} (c), the emergence of undamped and extra plasmon 
modes is a direct result of the SOI in the graphene system, which opens a EHC gap and separates 
the $\epsilon_r$ peak attributed to inter-band transitions from the $\epsilon_r$ dip 
attributed to intra-band transitions. The plasmon spectrum is a result of the interplay 
between the intra- and inter-band contributions to the dielectric function and may 
change significantly as parameters of the system such as the temperature and the electron 
density vary as discussed in the following.

By applying a gate voltage to a graphene layer, one can control the
electron density and the Fermi energy in the system \cite{novo,zhan}
and it is interesting to see the corresponding change in the excitations.
Due to the symmetry of the band structure, systems having the same
density of electrons or holes are equivalent. Here we consider a
system with a net electron density and positive Fermi energy $E_F$.
For a Dirac gas in a graphene without a SOI, the extra electrons in
the conduction band reduce the inter-band scattering rate but
enhances the intra-band scattering by increasing the length of the Fermi ring.
At $T=0$, a EHC gap of width $2v(k_F-q)$ is opened above the intra-band EHC edge 
in the range $0<q<k_F$. For $E_F > \Delta_{so}$ the $\epsilon_r$ has a dip in the 
higher energy side of $\omega=vq$ and a diminished peak in the lower energy side 
similar to the case of $E_F=0$ but for $T>0$ [Fig.~\ref{fig:fig1} (c)]. As for
the corresponding $\epsilon_i$, the peak in the higher energy side at $\omega=vq+0^+$ 
is flattened because the inter-band transition to the bottom of the conduction band 
is forbidden but a new peak appears in the other side at $\omega=vq-0^+$ due to the 
now allowed intra-band transition. As a result of the $\epsilon_r$ dip at 
$\omega=vq+0^+$ introduced by the intra-band transition, a
Landau damped plasmon mode appears near $\omega=vq$ and another one above,
with an approximate dispersion $\omega \propto \sqrt{q}$ at the small $q$. As compared to
the spectrum of an intrinsic graphene at a finite temperature [Fig.~\ref{fig:fig2} (a)], 
here the latter mode is undamped near $q=0$ and has a 
flatter dispersion slope near the crossing of the dispersion curve and the inter-band EHC 
edge. Instead of an emergence of the new plasmon modes when the plasmon 
dispersion curve exits the EHC gap of SOI the plasmon dispersion smoothly exits the 
EHC gap corresponding to a finite Fermi energy.
\begin{figure}
\begin{center}
\begin{picture}(450,160)
\put(-20,-30){\includegraphics{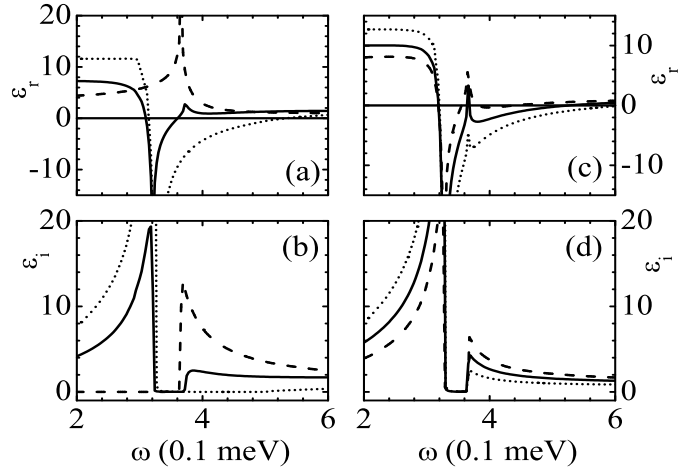}}
\end{picture}
\vspace{0cm} \protect\caption{Real ($\epsilon_r$) and imaginary
($\epsilon_i$) parts of the dielectric function vs $\omega$ at $T=0$
(left panels) and $T=2$ K (right panels) in a graphene of
$\Delta_{so}=0.08$ meV with $E_F=0$ (dashed), 0.25 meV (solid), and
0.4 meV (dotted). $q=0.05 \times 10^5$ cm$^{-1}$.} \label{fig:fig3}
\end{center}
\end{figure}

In the presence of the SOI, the physical scenario changes from the
$\Delta_{so}=0$ case for long-wavelength excitations if the
Fermi energy is located near the SOI gap. In Fig.~\ref{fig:fig3}, we
plot the dielectric function versus $\omega$ for different Fermi energies
at $T=0$ (left panels) and $T=2$ K (right panels) in a graphene with the
SOI strength of $\Delta_{so}=0.08$ meV. The dashed curves of Fig.~\ref{fig:fig1}
for the $E_F=0$ case is also plotted here for comparison.
At $T=0$, due to modification of the energy bands at
the bottom (or top), the intra-band EHC edge deviates from $\omega=vq$ to
a lower energy $\sqrt{\Delta_{so}^2+\hbar^2v^2(k_F+q)^2}
-\sqrt{\Delta_{so}^2+\hbar^2v^2k_F^2}$ for small $q$. In Fig.~\ref{fig:fig3}
(a) and (b), we observe a shift of the $\epsilon_r$ dip
and of the edge of the lower $\epsilon_i$ peak to an energy lower
than $vq$ for the $E_F=0.25$ meV curve (solid).
With the Fermi energy inside the conduction band,
the inter-band EHC edge at $T=0$ is expressed by $\omega_H$ when
$q>2k_F$ but shifts to a higher energy $\sqrt{\Delta_{so}^2+\hbar^2v^2k_F^2}+
\sqrt{\Delta_{so}^2+\hbar^2v^2(k_F-q)^2}$ for $q<2k_F$. As a result,  
the solid $\epsilon_r$ peak and the upper $\epsilon_i$ gap edge ($k_F=0.036$ cm$^{-1}$) 
have higher energies than the dashed ones ($k_F=0$) in Fig.~\ref{fig:fig3} (a) and (b).
When the Fermi energy
becomes higher, for example, as shown by the dotted curve of $E_F=0.4$ meV in
Fig.~\ref{fig:fig3} (a) and (b), the system reverts to the Dirac gas
and the dielectric function becomes similar to that in the
$\Delta_{so}=0$ case. From the solid curve for $\epsilon_r$ in Fig.~\ref{fig:fig3} 
(a) we observe that the SOI shifts the inter-band $\epsilon_r$ peak to a higher energy and
can change the weakly damped plasmon mode inside the inter-band EHC to
an undamped mode in the EHC gap created by the SOI.

\begin{figure}
\begin{center}
\begin{picture}(220,220)
\put(-15,25){\includegraphics{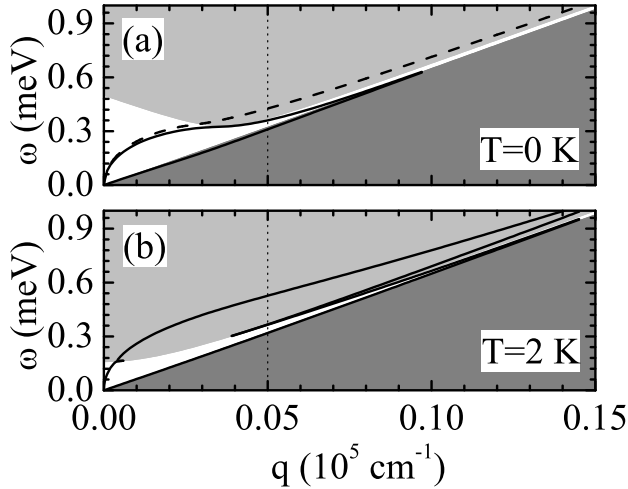}}
\end{picture}
\vspace{-1.4cm} \protect\caption{The same as in Fig.~\ref{fig:fig2} but
with $E_F=0.25$ meV at $T=0$ (a) and $T=2$ K (b). The solid curves are for 
$\Delta_{so}=0.08$ meV and the dashed in (a) for $\Delta_{so}=0$. The corresponding 
dielectric functions for the values of $q$
indicated by the vertical dotted lines are given in
Fig.~\ref{fig:fig3} by the solid curves.} \label{fig:fig4}
\end{center}
\end{figure}

At a finite temperature, however, the
restriction to single-particle transitions by the Fermi energy is
relaxed and both the intra- and inter-band EHC edges are the same as
those for $E_F=0$. In Fig.~\ref{fig:fig3} (c), where $\epsilon_r$ at
$T=2$ K is shown, the peaks and the dips of all the curves have the
same energy and  $\epsilon_i$ for different Fermi energies in
Fig.~\ref{fig:fig3} (d) have the same gap. These characteristics of
the single-particle excitation continuum edges are also shown in
Fig.~\ref{fig:fig4} by the light and dark shades for inter-
and intra-band EHC's respectively.

The effect of SOI on the zero-temperature plasmon spectrum of a graphene with 
a finite Fermi energy outlined in Fig.~\ref{fig:fig2} (a) is illustrated by its 
dispersion in Fig.~\ref{fig:fig4} (a). At a wavevector $q$ close to or higher than $k_F$, 
strong inter-band transitions between states of similar pseudospin $|\bm{k},+\rangle$ 
and $|\bm{k}',-\rangle$ with $\bm{k}$ anti-parallel to $\bm{k}'$ are possible. The 
contribution of these transitions at a finite $\Delta_{so}$ increases the real part of 
the dielectric function at high energies and moves one zero of $\epsilon_r$ to a 
lower energy. As a result, the corresponding plasmon dispersion curve is excluded 
from the inter-band EHC and confined in the EHC gap opened by the finite Fermi energy 
and the SOI. At a finite temperature $T=2K$, the electrons spread to higher energies
and the inter-band transition rate is reduced at large $q$ but increased at a small 
$q$. In this case, the dispersion curve of the mode above $vq$ remains almost intact 
except a splitting to three modes when it enters the inter-band EHC near $q \simeq 
0.004 \times 10^{5}$ cm$^{-1}$ as shown in Fig.~\ref{fig:fig4} (b). Being different from 
the behavior exhibited in Fig.~\ref{fig:fig2} (b), here the two extra plasmon modes along 
$\omega_H$ merge with each other soon after the splitting and emerge again at a 
higher $q$.

In summary, we have derived the electron-hole
propagator and the dynamically screened Coulomb interaction matrix of an
electron system in graphene with spin-orbit interaction
in the random phase approximation, using the standard
and widely used technique for multicomponent systems. In a system of
intrinsic graphene without the SOI, our result reduces to
the analytical result of a Dirac electron gas previously obtained by
the renormalization group theory. The spin-orbit interaction changes the
Dirac gas semimetal to a narrow gap semiconductor system. The carriers
far from the gap remain as the Dirac type while those near the gap shows 
different characteristics. At a finite temperature, the SOI splits the
inter-band single-particle continuum from the intra band one and
an undamped plasmon mode exists in the gap of the single-particle
excitation spectrum. For a gap of 0.16 meV, this plasmon mode exists
in a range of wave vector of the order of 0.1$\times 10^5$cm
$^{-1}$, which could perhaps be observed in experiments. With net electrons or holes 
in a graphene, a plasmon mode above the intra-band single-particle continuum exists 
and the Coulomb interaction is screened at a high energy even at the zero temperature. 
The SOI may push this mode to the single-particle continuum gap or split it into 
three modes. In contrast to the negligible effect of SOI to the dynamic Coulomb 
screening and the plasmon spectrum of a Fermi gas in a InGaAs quantum well \cite{wang}, 
the SOI may change significantly these properties of a Dirac gas in a graphene.

One of the authors (X.F.W) thanks P. Vasilopoulos for helpful discussions.
The work has been supported by the Canada Research Chair
Program and a Canadian Foundation for Innovation (CFI) Grant.

\end{document}